\documentclass[iop]{emulateapj}

\usepackage{amsmath}
\usepackage{epsfig}
\usepackage{graphicx}
\usepackage{apjfonts}
\usepackage{natbib}
\usepackage{psfrag}
\usepackage[colorlinks=true,citecolor=blue]{hyperref}
\newcommand{\sini}{\sin\,i}
\newcommand{\msini}{\langle\sin\,i\rangle}
\newcommand{\vsini}{V\!\sin\,i}
\newcommand{\mvsini}{\langle V\!\sin\,i\rangle}
\newcommand{\mv}{\langle V\rangle}
\newcommand{\mvm}{\langle V\rangle_{\rm CR}}

\newcommand{\momega}{\langle \Omega\rangle}
\newcommand{\momegam}{\langle \Omega\rangle_{\rm Model}}
\newcommand{\kms}{\,{\rm km\,s}^{-1}}

\newcommand{\rsun}{\,R_{\sun}}
\newcommand{\mrkic}{\tilde{R}_{\rm KIC}}
\newcommand{\rkic}{R_{\rm KIC}}
\newcommand{\mprot}{\tilde{P}_{\rm rot}}
\newcommand{\rest}{R_{\rm est}}
\newcommand{\tef}{T_{\rm eff}}
\newcommand{\prot}{P_{\rm rot}}

\slugcomment{}

\shorttitle{Chandrasekhar's relation and the stellar rotation in the {\it Kepler} field}
\shortauthors{Silva, Soares \& de Freitas}

\begin{document}

\title{Chandrasekhar's relation and the stellar rotation in the {\it Kepler} field}

\author{J. R. P. Silva, B. B. Soares}
\affil{Grupo de Atsroestat\'istica, Departamento de F\'{i}sica, Universidade do Estado do Rio 
Grande do Norte, Mossor\'o--RN, Brazil}
\email{joseronaldo@uern.br}

\and

\author{D. B. de Freitas}
\affil{Departamento de F\'{i}sica, Universidade Federal do Rio Grande do Norte, Natal--RN, Brazil}

\submitted{Received 2014 May 16; accepted 2014 September 26}

\begin{abstract}

According to the statistical law of large numbers, the expected mean of identically distributed 
random variables of a sample tends toward the actual mean as the sample increases.
Under this law, it is possible to test the Chandrasekhar's relation (CR), $\mv=(\pi/4)^{-1}\mvsini$, using
a large amount of $\vsini$ and $V$ data from different samples of similar stars. In this 
context, we conducted a statistical test to check the consistency of the CR in the {\it Kepler} 
field. In order to achieve this, we use three large samples of $V$ obtained from {\it Kepler} 
rotation periods and a homogeneous control sample of $\vsini$ to overcome the scarcity of $\vsini$ 
data for stars in the {\it Kepler} field. We used the bootstrap-resampling method to estimate the 
mean rotations ($\mv$ and $\mvsini$) and their corresponding confidence intervals for the stars 
segregated by effective temperature. Then, we compared the estimated means to check the consistency 
of CR, and analyzed the influence of the uncertainties in radii measurements, and possible selection 
effects. We found that the CR with $\msini=\pi/4$ is consistent with the behavior of the $\mv$ as a 
function of $\mvsini$ for stars from the {\it Kepler} field as there is a very good agreement 
between such a relation and the data.

\end{abstract}


\section{Introduction}
\label{introduction}

Studies on stellar rotation distribution are performed from either equatorial rotations, $V$, usually
calculated from rotation period measurements, or projected axial rotations $\vsini$, measured from the
spectral line profile broadening. When measurements of both $V$ and $\vsini$ are available, one can
fairly derive the inclination angle of the rotational axis to the line of sight. This makes it 
possible to determine the distribution of the inclination angles of the stars and hence the mean 
inclination angles of the rotational axes. In turn, the distribution of the inclination angles can 
provide evidence on the orientation of the rotational axes of a stellar population, and then 
determine whether or not these axes are randomly oriented.

The lack of observational evidence \citep{Struve45,Abt01,Jackson10}, in addition to the scarce works
pointing to a possible deviation from randomness \citep[eg.,][]{Vinketal05,Soares11}, which leads to a
modification of the assumption of the randomness\footnote{This hypothesis is widely ascribed 
to \citet{Chandrasekhar50}, however \citet{vandien48} and \citet{Struve45} suggested 
it earlier.} for the orientation of the stellar rotational axes, corroborate the frequency function 
of the inclination angles of the rotational axes with respect to the line of sight as $\sini$
\citep[cf.][]{vandien48,Chandrasekhar50}. A consequence of the assumption of the randomness is the widespread
relation between the mean rotational velocities obtained by \citet{Chandrasekhar50}: 
$\mv=(\pi/4)^{-1}\mvsini$.

It is worth mentioning that regardless of what the frequency function of the inclination angles is, 
it is always possible to obtain the relation $\mvsini=\mv\msini$, by assuming that the mean $V$ and 
$\sini$ are independent of each other. In fact, \citet{Chandrasekhar50} warned of the possibility  
that the inclination angle can have a frequency function other than $\sini$ and that relation 
between the mean averages remains valid. Much has been released by using the Chandrasekhar's 
ralation (CR), though a few works have been achieved to verify the validity of the frequency 
function $\sini$ and the constant $\pi/4$ \citep[eg.,][]{Abt01,Jackson10,Soares11}. Here we should 
mention \citet{Bernacca70} who stated that, when considering the rotation break-up limit, the 
inclination angle $i$ have a frequency function other than $\sini$.

Nowadays, new instruments and telescopes, such as {\it Kepler} \citep{Koch2010} and CoRoT
\citep{Baglin06}, can provide high-precision photometry for measurements of $V$ as much as $\vsini$
to make possible statistical tests to investigate the inclination angle distribution to
determine its true frequency function. Motivated by the plenty of observational data stemmed from these new
telescopes, we carried out the present work. Aiming to advance discussions about the validity of 
the average $\msini=\pi/4$, this work aimed to estimate the mean inclination angles of rotational 
axes for field dwarf stars while checking the CR for the stars in the {\it Kepler} field.

\section{Work sample}
\label{worksample}

The complete work sample consists of one control sample of $\vsini$ measurements and three samples 
of rotation period data for single main-sequence field stars with effective temperature, 
$\tef$, between 4\,300 and 7\,000\,K (see Section \ref{temperatures}). The control sample, 
hereafter N04, is from the magnitude-complete and kinematically unbiased sample of stars 
released by \citet{Nordstrom04}. The rotation periods, $\prot$, were determined in three studies 
using light curves (LC) from the {\it Kepler} mission archive. The samples contain rotation 
periods that are referred to R13, N13, and M14. Next, we will describe the main 
characteristics of the samples and the methods used in their measurements. We refer the reader to 
the original sources for complete details.

\subsection{Projected Equatorial Rotations}
\label{vsini}

Measurements of $\vsini$ were carried out with the CORAVEL spectrometer \citep{Baranne79, Benz81} using the 
calibration technique by \citet{Benz84}. Such a technique allows the separation of 
the observational signature of the rotation from the micro and macroturbulence, pressure, and 
Zeeman splitting, which permits measurements of $\vsini$ for slow rotators with an 
accuracy of $1\kms$. For high rotators, the shallow cross-correlation profiles decrease the 
accuracy of the $\vsini$ measurements. Therefore, the $\vsini$ values higher than $30\kms$ are 
given with uncertainties of $5-10\kms$. To compose the control sample, we selected all of the 
stars with $\vsini$ data in \citet{Nordstrom04}, except the objects classified as binary or cluster 
stars \citep[see also][]{Holmberg09}. The control sample has 7\,542 $\vsini$ data.

\subsection{Rotation Periods}
\label{periods}

The R13 sample is composed of 20,631 $\prot$ measured by \cite{Reinhold13} using data in the 
Quarter 3 (Q3) processed with the PDC-MAP pipeline. The stars were selected according to the 
criteria $\log g > 3.5$ and variability amplitude $R_{\rm{var}}\geq 0.003$. The parameter 
$R_{\rm{var}}$ was defined as the 5th and 95th percentile of normalize flux, with 4\,h boxcar 
smoothed differential LC. To detect periods, they used the Generalized Lomb--Scargle periodogram 
\citep{Zechmeister09} to LC in 2\,h bins. To minimize the aliasing caused by active regions 
on opposite hemispheres of the star, they compared the first two periods from the global sine fit. 
When the difference between the first and second period was less than 5\%, they selected the longer 
one as the stellar rotation period.  Aiming to exclude the pulsators, they selected a lower limit 
period equal to 0.5\,day. A 45\,day upper limit for the period was also selected to avoid 
instrumental effects, visible on timescales of the Q3 duration. In addition, all the periods close 
to the orbital period of known binaries were excluded.

Rotation periods in M14 are from \citet{McQuillan14}, and have 27,631 stars. Those authors used 
the LC from Q3 to Q14, which were corrected for instrumental systematics using PDC-MAP. They used 
the $\tef$-$\log g$ and color-color cuts by \citet{Ciardi11} to select only main-sequence stars. 
The rotation periods were obtained using the method described by \citet{McQuillan13}, which is based 
on the autocorrelation function (ACF) of the LC. Such a method consists of measuring the degree of 
self-similarity of the LCs over a range of lags. The repeated spot-crossing signatures lead to ACF 
peaks at lags corresponding to $\prot$ and their integer multiples, in the case of rotational 
modulation. The authors considered only periods between 0.2 and 70\,days to avoid excess of false 
positives from pulsators, and long periods difficult to determine. In addition, they selected only 
late-F stars ($\tef\lesssim6\,500$\,K\footnote{The effective temperatures adopted by 
\citet{McQuillan14} were calculated from the dereddened color index $\bv$ using the calibration by 
\citet{Sekiguchi00}.} because they were interested in the stars with convective envelopes. The 
eclipsing binaries and {\it Kepler} Objects of Interest (KOIs) were removed when identified.

The N13 sample contains 10,426 stars with the rotation period measured by \citet{Nielsen13}. These 
authors used eight quarters, from Q2 to Q9, corrected with the PDC-MAP. They applied a 
Lomb--Scargle periodogram \citep{Frandsen95} to determine the rotation periods from the LC. In their 
study, only active stars with $\log(g)\geq3.4$ were selected, and a periodogram peak at least four 
times the white noise, estimated using the root mean square (rms) of the time series. The rotation 
period was determined as the median of the stable periods over the quarters. With the objective to 
avoid pulsators, they selected only $\prot\geq1$\,day. An upper limit period was assumed as 
30\,days because their method depends on the stable periods and does not work for longer periods. 
The objects known as eclipsing binaries, and KOIs were excluded from the sample.

The median of the rotation periods segregated by interval of effective temperature for the samples R13, M14,
and N13 are given in Table \ref{data}.

\begin{deluxetable*}{@{\extracolsep{-1.5mm}}p{15mm}r|rr|rrrrr|rrrrr|rrrrr}
\tabletypesize{\footnotesize}
\tablewidth{18cm}
\tablecaption{Main Parameters of the Stars Analyzed in this Work \label{data}}
\tablehead{
 \multicolumn{2}{c|}{\nodata}  &\multicolumn{2}{c|}{N04}  &\multicolumn{5}{c|}{R13} &
\multicolumn{5}{c|}{M14}  &\multicolumn{5}{c}{N13}\\[0.1ex]
\cline{1-19}\\[-1ex]
 \multicolumn{1}{c}{$\tef$} &  \multicolumn{1}{c|}{$\rest$}   &  \multicolumn{1}{c}{$N$}  &
\multicolumn{1}{c|}{$\mvsini$}
&  \multicolumn{1}{c}{$N$}  &  \multicolumn{1}{c}{$\mv$}    &  \multicolumn{1}{c}{$\momega$}  &
 \multicolumn{1}{c}{$\mrkic$} & \multicolumn{1}{c|}{$\mprot$} &  \multicolumn{1}{c}{$N$}  &
\multicolumn{1}{c}{$\mv$} & \multicolumn{1}{c}{$\momega$} &  \multicolumn{1}{c}{$\mrkic$} &
\multicolumn{1}{c|}{$\mprot$}
&  \multicolumn{1}{c}{$N$}  &  \multicolumn{1}{c}{$\mv$}    &  \multicolumn{1}{c}{$\momega$}  &
 \multicolumn{1}{c}{$\mrkic$} &  \multicolumn{1}{c}{$\mprot$}
}
\startdata
$ 4300 - 4600 $ & 0.66  & 17 & $3_{- 0.8 }^{ + 0.9 }$ & 1962 & $2.7_{- 0.2 }^{ + 0.2
}$ &$ 3.8_{- 0.3 }^{+ 0.3}$ & 0.68 & 19.8 & 2383 & $2.7_{- 0.3 }^{ + 0.3 }$ &$ 3.8_{- 0.4 }^{+
0.5 }$ & 0.68 & 23.7 & 652 & $3.5_{- 0.3}^{ + 0.3 }$ &$ 4.8_{- 0.4 }^{+ 0.4 }$ & 0.70 & 14.0
\\[0.4ex]
$ 4600 - 4800 $ & 0.69 & 35 & $2.4_{- 0.6 }^{ + 0.6 }$ & 1205 & $3.5_{- 0.3 }^{ + 0.4
}$ &$ 3.8_{- 0.4 }^{+0.4 }$ & 0.75 & 19.8 & 1365 & $3.4_{- 0.4 }^{ + 0.5 }$ &$ 3.6_{- 0.5 }^{+ 0.5
}$ & 0.75 & 24.0 & 394 & $4.8_{-0.5 }^{ + 0.5 }$ &$ 4.9_{- 0.5 }^{+ 0.6 }$ & 0.77 & 14.0 \\[0.4ex]
$ 4800 - 5000 $ & 0.72 & 124 & $2.7_{- 0.4 }^{ + 0.5 }$ & 1731 & $4.1_{- 0.4 }^{ + 0.4
}$ &$ 4.1_{- 0.3 }^{+0.3 }$ & 0.91 & 19.6 & 1751 & $3.8_{- 0.5 }^{ + 0.5 }$ &$ 4.2_{- 0.5 }^{+ 0.6
}$ & 0.88 & 23.5 & 597 & $5.8_{-0.7 }^{ + 0.9 }$ &$ 5.5_{- 0.5 }^{+ 0.5 }$ & 0.92 & 13.6 \\[1.4ex]
$ 5000 - 5200 $ & 0.76 & 236 & $2.8_{- 0.4 }^{ + 0.4 }$ & 2197 & $4.2_{- 0.4 }^{ + 0.4
}$ &$ 4.0_{- 0.3 }^{+0.3 }$ & 0.90 & 19.1 & 2410 & $4.1_{- 0.5 }^{ + 0.5 }$ &$ 4.2_{- 0.4 }^{+ 0.5
}$ & 0.89 & 22. & 772 & $6.4_{-0.7 }^{ + 0.8 }$ &$ 5.6_{- 0.4 }^{+ 0.4 }$ & 0.91 & 13.1 \\[0.4ex]
$ 5200 - 5400 $ & 0.81 & 300 & $2.9_{- 0.3 }^{ + 0.3 }$ & 2511 & $4.8_{- 0.4 }^{ + 0
}$ &$ 4.8_{- 0.3 }^{+ 0.3}$ & 0.86 & 17.6 & 2804 & $5.2_{- 0.5 }^{ + 0 }$ &$ 5.6_{- 0.6 }^{+ 0.6
}$ & 0.86 & 20.3 & 1006 & $5.8_{- 0.4}^{ + 0 }$ &$ 5.7_{- 0.3 }^{+ 0.4 }$ & 0.89 & 12.1 \\[0.4ex]
$ 5400 - 5600 $ & 0.88 & 682 & $3.3_{- 0.3 }^{ + 0.3 }$ & 2842 & $4.5_{- 0.3 }^{ + 0
}$ &$ 4.7_{- 0.2 }^{+ 0.3}$ & 0.87 & 16.2 & 3216 & $5.2_{- 0.6 }^{ + 0 }$ &$ 5.5_{- 0.4 }^{+ 0.5
}$ & 0.86 & 18.5 & 1129 & $5.7_{- 0.3}^{ + 0 }$ &$ 5.8_{- 0.3 }^{+ 0.3 }$ & 0.87 & 11.6 \\[1.4ex]
$ 5600 - 5800 $ & 0.96 & 1010 & $3.6_{- 0.2 }^{ + 0.3 }$ & 2552 & $5.1_{- 0.3 }^{ +
0.4 }$ &$ 5.1_{- 0.3 }^{+0.3 }$ &0.87 & 15.3 & 3237 & $6.0_{- 0.6 }^{ + 0.6 }$ &$ 6.2_{- 0.6 }^{+
0.6 }$ & 0.87 & 17.5 & 1228 & $6.2_{-0.4 }^{ + 0.4 }$ &$ 6.2_{- 0.3 }^{+ 0.3 }$ & 0.88 &11.0
\\[0.4ex]
$ 5800 - 6000 $ & 1.05 & 1253 & $5.4_{- 0.3 }^{ + 0.3 }$ & 2107 & $6.8_{- 0.5 }^{ +
0.5 }$ &$ 6.5_{- 0.4 }^{+0.4 }$ &0.93 & 12.1 & 2930 & $7.2_{- 0.6 }^{ + 0.6 }$ &$ 6.9_{- 0.5 }^{+
0.5 }$ & 0.94 & 14.6 & 1138 & $7.7_{-0.5 }^{ + 0.6 }$ &$ 7.2_{- 0.4 }^{+ 0.4 }$ & 0.95 & 9.9
\\[0.4ex]
$ 6000 - 6200 $ & 1.14 & 1416 & $9_{- 0.4 }^{ + 0.4 }$ & 1817 & $8.4_{- 0.5 }^{ + 0.5
}$ &$ 7.9_{- 0.4 }^{+0.4 }$ & 0.99 & 9.7 & 3042 & $9.3_{- 0.7 }^{ + 0.8 }$ &$ 8.4_{- 0.6 }^{+ 0.6
}$ & 1.01 & 11.9 & 1318 & $9.3_{-0.6 }^{ + 0.6 }$ &$ 8.2_{- 0.4 }^{+ 0.5 }$ & 1.01 & 9.1 \\[1.4ex]
$ 6200 - 6400 $ & 1.24 & 1278 & $14.7_{- 0.6 }^{ + 0.6 }$ & 987 & $13.9_{- 1.1 }^{ +
1.2 }$ &$ 12.0_{- 0.8}^{+ 0.8 }$ &1.04 & 6.6 & 2300 & $13.7_{- 0.9 }^{ + 1 }$ &$ 11.1_{- 0.7 }^{+
0.7 }$ & 1.08 & 8.5 & 966 &$13.5_{- 1 }^{ + 1 }$ &$ 11.1_{- 0.6 }^{+ 0.6 }$ & 1.06 & 7.0 \\[0.4ex]
$ 6400 - 6600 $ & 1.34 & 739 & $21.1_{- 0.9 }^{ + 0.9 }$ & 405 & $25.9_{- 2.5 }^{ +
2.5 }$ &$ 20.4_{- 1.7 }^{+1.8 }$ &1.19 & 3.5 & 1520 & $23.1_{- 1.5 }^{ + 1.7 }$ &$ 17.7_{- 1.0
}^{+ 1.2 }$ & 1.23 & 4.5 & 627 & $20.9_{-1.4 }^{ + 1.4 }$ &$ 16.0_{- 1.0 }^{+ 1.0 }$ & 1.24 & 4.1
\\[0.4ex]
$ 6600 - 6800 $ & 1.42 & 332 & $43.2_{- 2.9 }^{ + 3 }$ & 177 & $44.5_{- 5.9 }^{ + 6.1
}$ &$ 32.2_{- 3.5 }^{+3.5 }$ & 1.32 & 2.0 & 657 & $34.4_{- 2.7 }^{ + 2.9 }$ &$ 24.2_{- 1.8 }^{+
1.9 }$ & 1.35 & 3.0 & 348 & $28.6_{-2.4}^{ + 2.5 }$ &$ 20.2_{- 1.4 }^{+ 1.5 }$ & 1.39 & 2.9
\\[1.4ex]
$ 6800 - 7000 $ & 1.49 & 120 & $51.5_{- 5.1 }^{ + 5.5 }$ & 138 & $69.4_{- 6.8 }^{ +
7.6 }$ &$ 46.7_{- 4.6 }^{+5.0 }$ &1.51 & 1.1 & 16 & $43.6_{- 25.6 }^{ + 36.8 }$ &$ 23.0_{- 12.8
}^{+ 15.7 }$ & 1.58 & 5.2 & 251 &$38.8_{- 3.3 }^{ + 3.3 }$ &$ 24.5_{- 1.9 }^{+ 1.8 }$ & 1.51 & 2.1
\\[-1ex]
\enddata
\tablecomments{The errors in the mean rotations represent the bootstrapped 95\% confidence 
interval of the mean. The effective temperature is give in kelvins, estimated and median radii are 
given in $\rsun$, mean rotations $\mvsini$, and $\mv$ are given in $\kms$, $\momega$ is given in 
s$^{-1}$, and the median rotation period is given in days. The first Column presents the
effective temperature ranges. Second and third column the radii, $\rest$, estimated in the middle 
of 
the $\tef$ range using the data give in Table 1B by \citet{Gray92}. Third and forth columns present 
the number of rotators in the sample N04, and the mean rotations, $\mvsini$. Next columns show the 
parameters for the samples R13, N13, and M14 as indicated. These columns show the number of 
rotators, the mean rotations $\mv$, mean angular rotation, $\momega$, the median KIC radii, 
$\mrkic$, and the median rotation periods, $\mprot$, respectively}
\end{deluxetable*}

\subsection{Effective Temperatures}
\label{temperatures}

Effective temperatures of the stars in the N04 sample were determined by \citet{Nordstrom04} using 
the calibration by \citet{Alonso96}, and the dereddened color indices $b-y$, $c_1$, and $m_1$. These 
stars have $\tef$ in the range $4\,300$-$7\,000$\,K. Such a range defined the temperature interval 
of the stellar samples analyzed in this paper. Adopting a homogeneous scale of effective 
temperature for the stars in the {\it Kepler} field prevents the introduction of systematic bias as 
a result of using $\tef$ from different studies. Accordingly, the effective temperature of the {\it 
Kepler} targets was determined according to the homogeneous effective temperature scale by 
\citet{Pinsonneault12}. These authors used the Sloan Digital Sky Survey {\it griz} filters, tied to 
the fundamental temperature scale, to revise $\tef$ in the {\it Kepler} Input Catalog (KIC). As a 
result, they found a mean shift of about 215\,K toward higher temperatures relative to the KIC 
$\tef$. In the present work, a systematic shift to higher $\tef$ results in a systematic shift down 
in the mean rotations $\mv$ due to the displacement of some lower rotators to the next $\tef$ bin.

\section{Mean rotations and confidence-intervals}
\label{meanrotations}

Stellar rotation is correlated with spectral type or, equivalently, with effective temperature 
\citep[see Figure 2 in][]{Slettebak70}. Thus, to minimize the bias due to the mix of 
spectral types in our results, we binned the samples into 200\,K bins (except the first one, which 
has a bin width of 300\,K). Such a bin is the minimum range allowing a reasonable amount of stars 
($\gtrsim20$) for performing the bootstrap resampling.

True equatorial rotations (in $\kms$) were calculated using the equation $V=50.58\rkic\prot^{-1}$, 
where $\rkic$ is the KIC radius (in $\rsun$), and $\prot$ given in days. The angular rotations 
were calculated from $\prot$ according to the equation $\Omega=V\rkic^{-1}=50.58\prot^{-1}$, where 
$\prot$ is given in days, and $\Omega$ given in s$^{-1}$. The mean rotations, and their 95\% 
confidence intervals in each $\tef$ bin, were estimated using bootstrap resampling \citep{Efron87, 
Efron93}. The bootstrap-resampling method consists of generating a large number of data sets, each 
with an equal amount of data randomly drawn from the original data \citep[cf.][]{Feigelson12}. The 
way by which we proceeded was as follows. First, we performed a set of 1\,000 bootstrap replications 
of the mean rotation in the $\tef$ bin. The mean value of the distribution of these bootstrapped 
means was assumed to be the mean rotation in the $\tef$ bin. Then, we ranked the bootstrapped means, 
from the lower to the higher value, and took the 25th and the 975th means in the rank as the 
lower and upper limits of the confidence interval, respectively. The mean rotations and their 
confidence intervals in each $\tef$ bin are presented in Table \ref{data}.

\section{Some assumptions and approximations}
\label{assumption}

Before proceeding further, it is important to make clear our assumptions. These are only 
approximations of the reality, once we are neither dealing with unbiased samples, nor accounting 
for all the effects contributing to the observed signature of the rotation periods.

For determining \emph{exactly} the mean $\langle\sin{i}\rangle$ from $\langle
V\sin{i}\rangle/\langle V\rangle$ it is necessary to have the mean equatorial rotations and the 
mean projected rotational velocities, both from the same sample of stars. Unfortunately, it is very 
difficult to produce a significant sample of $V$ and $V\sin{i}$ measurements. Given that there is no 
privileged direction in the Galaxy for stellar rotation, or, in other words, any representative 
sample of $V$ or $V\sin{i}$ in any direction of the Galaxy should have similar behavior to another 
representative sample in a different direction, it is possible to estimate the mean 
$\langle\sin{i}\rangle$ from these representative samples. Since the sample is homogeneous and the 
data are independent of each other, the mean rotation for each sample should represent the mean 
stellar rotation in the Galaxy---according to the statistical law of large numbers which states 
that the sample mean converges to the distribution mean as the sample size increases. Thus, as the 
mean $\langle V\rangle$ (or $\langle V\sin{i}\rangle$) of a sample represents the $\langle V\rangle$ 
(or $\langle V\sin{i}\rangle$) in the Galaxy, it must also represent the mean of the same physical 
quantity of another significant sample of stars. Based on these considerations, we assumed that the 
mean $\langle V\sin{i}\rangle$ from N04 represents the mean $\langle V\sin{i}\rangle$ of the {\it 
Kepler} field samples, since the former represents the mean of the physical quantity of all the 
stars with $\tef$ in the considered temperature bin. Likewise, the mean $\langle V\rangle$ from the 
{\it Kepler} field samples---because of their significances (large number of stars)---must 
represent the mean $\langle V\rangle$ of the field stars whatever the direction of Galaxy; 
therefore, it also represents the mean $\langle V\rangle$ of the N04 sample.

We presumed that each bootstrapped confidence interval contains a value close to the mean 
rotation of the entire population of rotators with $\tef$ ranging in the considered temperature  
bin. This assumption is reasonable because, except in a very few cases, the $\tef$ bins contain a 
large amount ($\gtrsim120$) of stars with approximately the same spectral type. In addition, 
according to the Bootstrap Central Limit Theorem \citep{Singh81}, bootstrapped means tend to become 
closer to the population true mean as the number of bootstrap resampling increases. This is clearly 
only an approximation because, apart from the bias introduced by the methods in determining $\prot$ 
(or $\vsini$), there are additional biases in the samples. For instance, the stars in KIC 
are biased to solar-type stars, as the {\it Kepler} targets were selected in order to maximize the 
probability of detecting a planetary transit around stars similar to the Sun in mass and $\tef$ 
\citep[][see Figure 1]{Brown11, Reinhold13}. In addition, there is not sufficient information to 
remove all the binary stars or multiple systems from the samples.

We assumed rigid body rotation. In fact, as outlined by \citet{Reinhold13}, $\prot$ results from 
active regions with different velocities manifesting themselves as a superposition of different 
periods in the LC. They analyzed the distributions of $\prot$ and another period ($P'$), within 30\% 
of $\prot$, and found out in the remaining sines of their LC's that these distributions are similar, 
presenting mean values of $\prot=16.3\pm10.1$\,days, and $P'=13.3\pm7.3$\,days. They also found 
that the difference between the angular velocity at the stellar equator and at the pole increases 
weakly with the temperature in the range $3\,500$-$6\,000$\,K, which is nearly the temperature 
range we are considering. According to these results, the differential rotation does not seem to 
strongly affect our conclusions.

We also assumed that the rotation period of the stellar chromosphere, which yields $\prot$, is equal to the
rotation period of the photosphere, which yields $\vsini$. This assumption is concerned with the problem in
determining $\mv$ from $\mvsini$, highlighted by \citet{Soderblom85}. As this author pointed out, it is likely
that the stellar chromosphere rotates more rapidly than the photosphere as occur on the Sun.
Furthermore, \citet{Soderblom85} drew attention to the fact that the problem of the spatial orientation
of the rotation axes had two parts. First, the angle $i$, between the rotational axis and the line of sight,
and second, the position angle of that axis on the plane of the sky. In this respect, we 
make clear that we are concerned only with the first part.

Finally, we take for granted that $V$ and $\sini$ are independent variables, allowing us to equate
their means as $\mvsini=\mv\msini$.

\section{Results and discussion}
\label{res_disc}

\begin{figure*}
  \centering
  \includegraphics[width=\hsize]{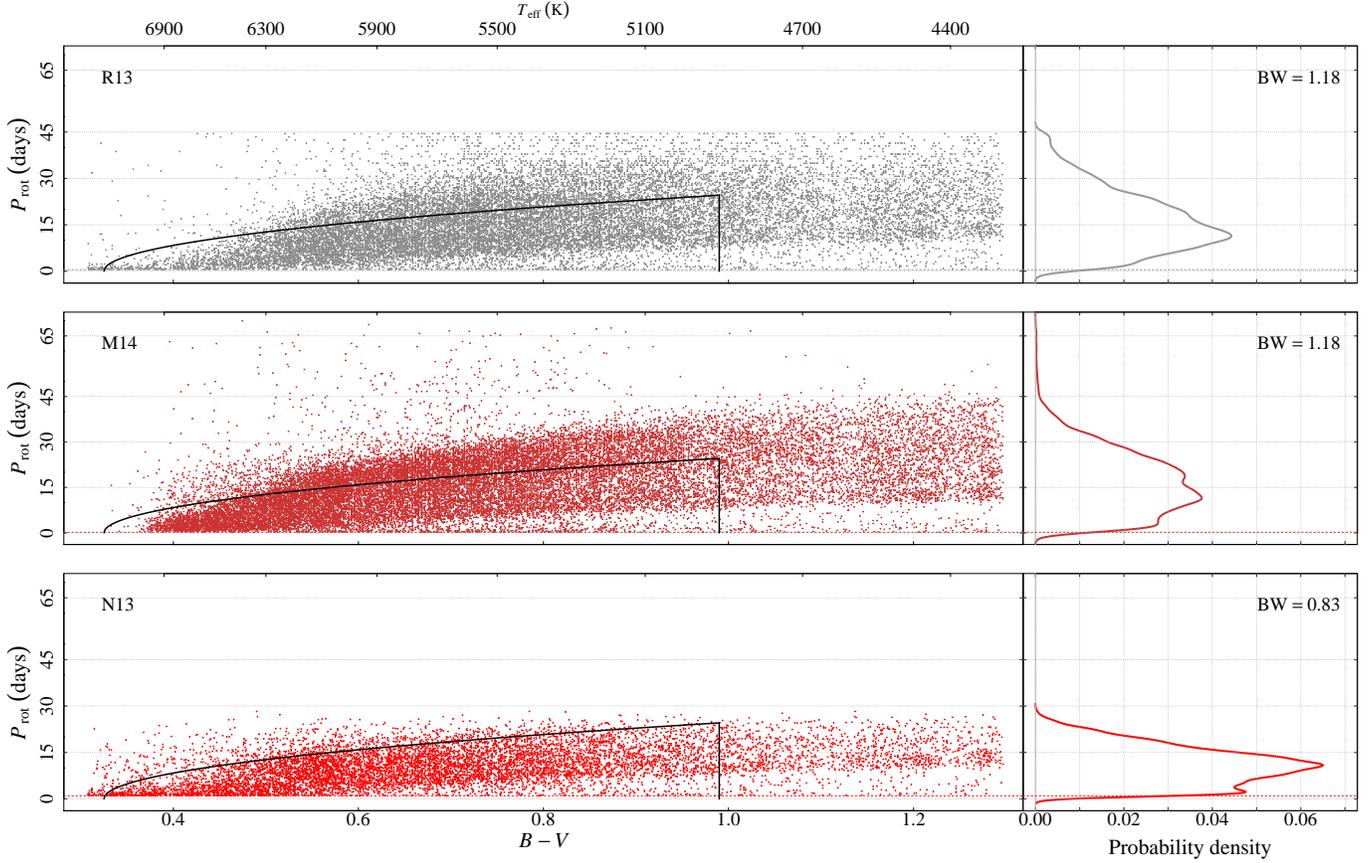}
  \caption{Left panels display the distribution of rotation periods, $\prot$, as a function of 
the color-index, \bv, for the sample R13, M14, and N13 as indicated. The regions limited by 
the curves in black enclose the stars with ages less than 2\,Gyr and temperature in the range of 
4\,900-7\,000\,K. The top axis corresponds to the effective temperature. Right panels show the 
kernel density estimates for $\prot$ for the distribution in the same color. The smoothing 
bandwidth, BW, is indicated in each panel.}
   \label{density}
 \end{figure*}

Samples R13, M14, and N13 are very different from each other because the original samples have different
selection criteria and methods in measuring $\prot$. About 70\% of the objects in M14 are not in 
N13. More than 37\% of the objects in the M14 sample are not in the R13 sample, and around 
23\% of the stars in the N13 sample do not belong to the R13 sample. The left panels of 
Figure \ref{density} present the distribution of $\prot$ as a function of the $\bv$ color index, 
estimated using a calibration between $\tef$ and $\bv$ obtained from the data give in Table 1B by 
\citet{Gray92}. The right panels shows the kernel density estimates of $\prot$. Although low and 
high rotators are present at all temperatures, it seems to be clear that there is a tendency for 
increasing rotation period with decreasing temperature, as predicted by the magnetic braking theory 
\citep{Schatzman62}. There is a peak of probability around $\prot=12$\,days, which for a star with 
radius $\sim1\rsun$ corresponds to a rotation of $V\sim4\kms$. A similar peak is observed in the 
distribution of $\vsini$ in Figure 4 by \citet{Nordstrom04}, where we see a peak at 
$\vsini\sim3$-$4\kms$. The periods in the R13 sample are limited to 45\,days; however, there are 
only a few rotators with higher periods, which we can also see in the middle panel, for M14. 
The M14 sample presents a few stars with $\tef\gtrsim6\,800$\,K, because the stars 
are restricted to spectral region late-F \citep{McQuillan14}. We also observe that the N13 sample 
is strongly biased against long periods, as we can observe by the sharp decline in the probability 
density function for the longer periods. This is due a limitation of the method used by 
\citet{Nielsen13}, as mentioned in Section \ref{periods}. Finally, the sample N13 presents a sharp 
fall in the probability density function for the shorter periods, as a result of their lower limit 
period of one day.
 
\begin{figure*}
 \centering
 \includegraphics[width=17cm]{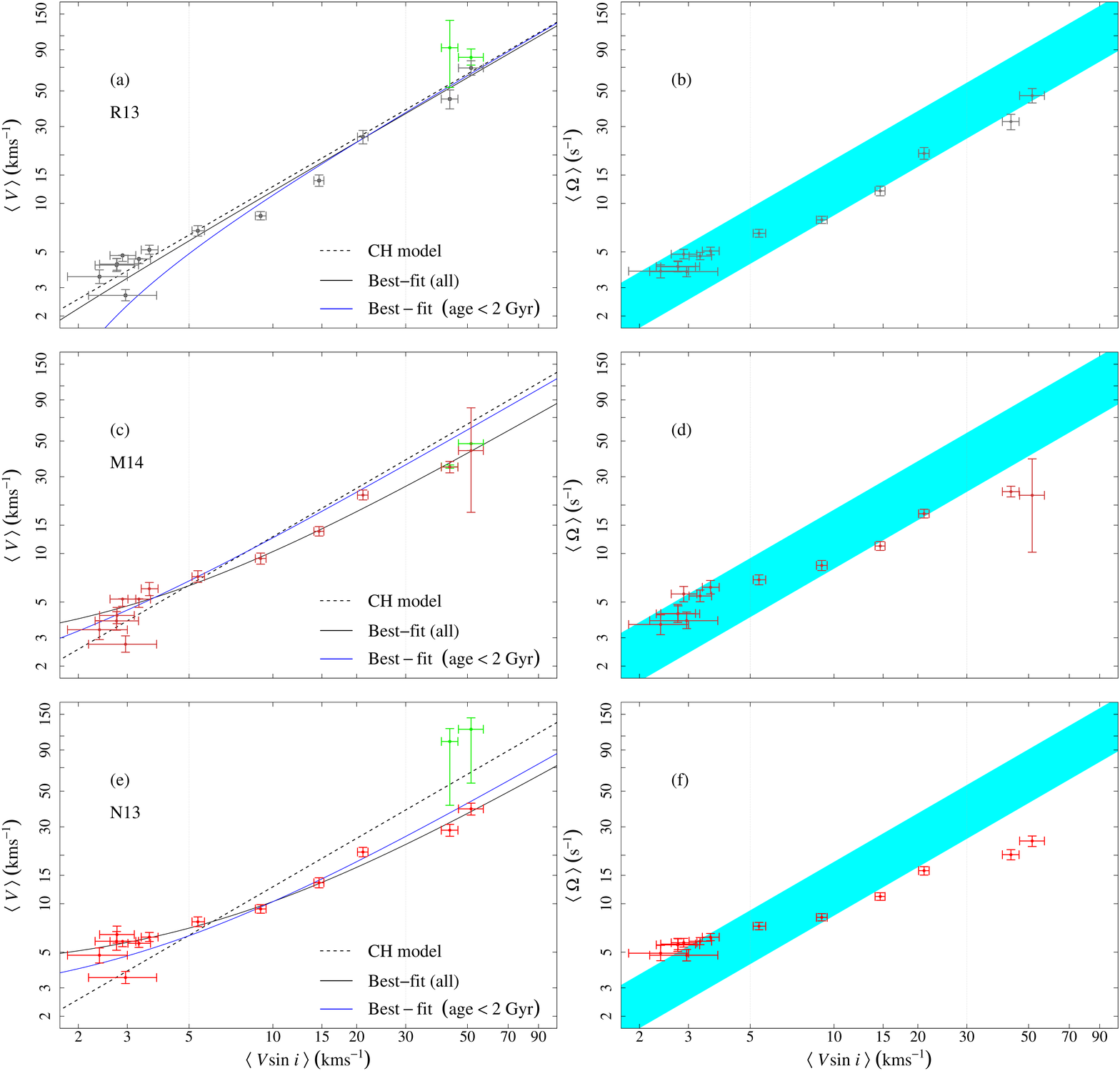}
 \caption{\small Left-side panels show the mean true rotations as a function of the mean projected rotations. 
The error bars are bootstrapped 95\% confidence intervals. The dashed line is the Chandrasekhar's 
relation, the continuous black line is the best-fit line for the data, and the blue line is the 
best-fit line for the data of the subsample containing only stars with ages $<2$\,Gyr and 
$4\,900\leqq \tef\leqq7\,000$\,K. The green dots represent the corrected means from the NPMLE (see 
the text), and their correspondent error bars are the 95\% confidence bands of the NPMLE estimated 
through bootstrap resamples with 1000 replacements. The right-side panels present the mean true 
angular rotation as a function of the mean projected rotations. The cyan region corresponds to 
the mean angular rotation calculated from $\mvsini$ by $\momega=(R\,\pi/4)^{-1}\mvsini$, where radii $R$ span 
in a range from the minimum to the maximum median KIC radii, $\mrkic$, given in Table \ref{data}. The axes are 
in logarithmic scale and the error bars represent the bootstrapped 95\% (percentile) confidence 
interval for the means.}
 \label{means}
\end{figure*}

Figure \ref{means} displays the distribution of $\mv$ as a function of $\mvsini$, in logarithmic 
scale. The $\mvsini$ data from N04 spans over the horizontal axes, and the vertical axes 
correspond to $\mv$ (left-side panels), or $\momega$ (right-side panels). First, we will analyze the 
left panels, then we will examine the panels on the right. In the left-side panels, the dashed line 
represent the CR model $\mv=(\pi/4)^{-1}\mvsini$, and the continuous line is the 
best-fit\footnote{The best-fit line was obtained using linear regression analysis, and the least 
squares parameter was obtained using the QR decomposition method.} for the data. There is a very 
good agreement between the data from R13 and the CR model as can be observed comparing the CR model 
with the best fit line in panel (a). The best-fit line is given by $\mv=1.2\mvsini-0.2$, with the 
rms of the best fit ${\rm rms_f}=3.2\kms$. The rms of the differences between $\mv$ and the CR model 
is ${\rm rms_{CR}}=3.5\kms$. The model is accepted with a 95\% confidence level by six of 13 
groups. Table \ref{disc} presents the median, minimum and maximum difference between $\mvm$ and 
$\mv$, relatively to $\mv$, for low, moderate, and high rotators, namely groups with 
$\mvsini\lesssim5\kms$, $5\lesssim\mvsini\lesssim30\kms$, and $\vsini\gtrsim30\kms$, respectively. 
The largest median discrepancy occurs to the groups with moderate rotators. However, even such a 
discrepancy can be explained by considering the uncertainties in the KIC radii, as we will 
see below.

\begin{table}
\caption{Relative Differences between the Mean Rotation Estimated with the CR Model, $\mvm$, and 
the Mean Rotation Calculated using $\prot$ and $\rkic$, $\mv$.
\label{disc}}
\centering
\begin{tabular}{|r|r|rrr|}
\tableline\tableline
\multicolumn{1}{|c|}{Sample} & \multicolumn{1}{c|}{$\mvsini$} &
\multicolumn{1}{r}{$\delta_{\rm min}$} & \multicolumn{1}{c}{$\delta_{\rm med}$}
& \multicolumn{1}{r|}{$\delta_{\rm max}$}\\
\multicolumn{1}{|c|}{} & \multicolumn{1}{c|}{($\kms$)} & \multicolumn{1}{c}{(\%)} &
\multicolumn{1}{c}{(\%)}& \multicolumn{1}{c|}{(\%)}\\
\tableline
    & $\lesssim5$    & 9 & 12 & 36\\
R13 & $\sim5$-30 & 1 & 27 & 30\\
    & $\gtrsim30$    & 3 & 4  & 18\\\hline
    & $\lesssim5$    & 6 & 15 & 28\\
M14 & $\sim5$-30 & 4 & 19 & 28\\
    & $\gtrsim30$    &12 & 40 & 47\\\hline
    & $\lesssim5$    & 6 & 29 & 36\\
N13 & $\sim5$-30 & 9 & 19 & 30\\
    & $\gtrsim30$    &22 & 55 & 73\\
\tableline
\end{tabular}
\tablecomments{The median, minimum, and maximum 
of the differences are namely $\delta_{\rm med}$, $\delta_{\rm min}$, and $\delta_{\rm max}$, 
respectively.}
\end{table}

The agreement between the CR model and the data for the M14 sample is not as good as for the  
R13 sample. However, there is reasonable consistency $\mvm$ and $\mv$ as we can see in panel 
(b). The best-fit line follows the equation $\mv=0.8\mvsini+2.3$, with ${\rm rms_f}=1.5\kms$, and 
${\rm rms_{CR}}=8.6\kms$. According to the best-fit line, the groups with low rotators tend to 
present $\mv>\mvm$. The groups with moderate and high rotators tend to have $\mv<\mvm$, and there is 
even a group presenting $\mvsini>\mv$. A similar behavior is presented by the N13 sample, as we 
see in panel (e). The best-fit line for N13 data follows the equation $\mv=0.7\mvsini+3.8$, with 
${\rm rms_f}=1.6\kms$, and ${\rm rms_{CR}}=10.8\kms$. The median relative differences between $\mv$ 
and $\mvm$ for the samples M14 and N13, are shown given in Table \ref{disc}. These discrepancies 
seem to be due primarily to two factors, the uncertainties in the KIC radii, and selection effects 
in the samples, as we will discuss next.

\begin{figure}
  \centering
  \includegraphics[scale=0.55]{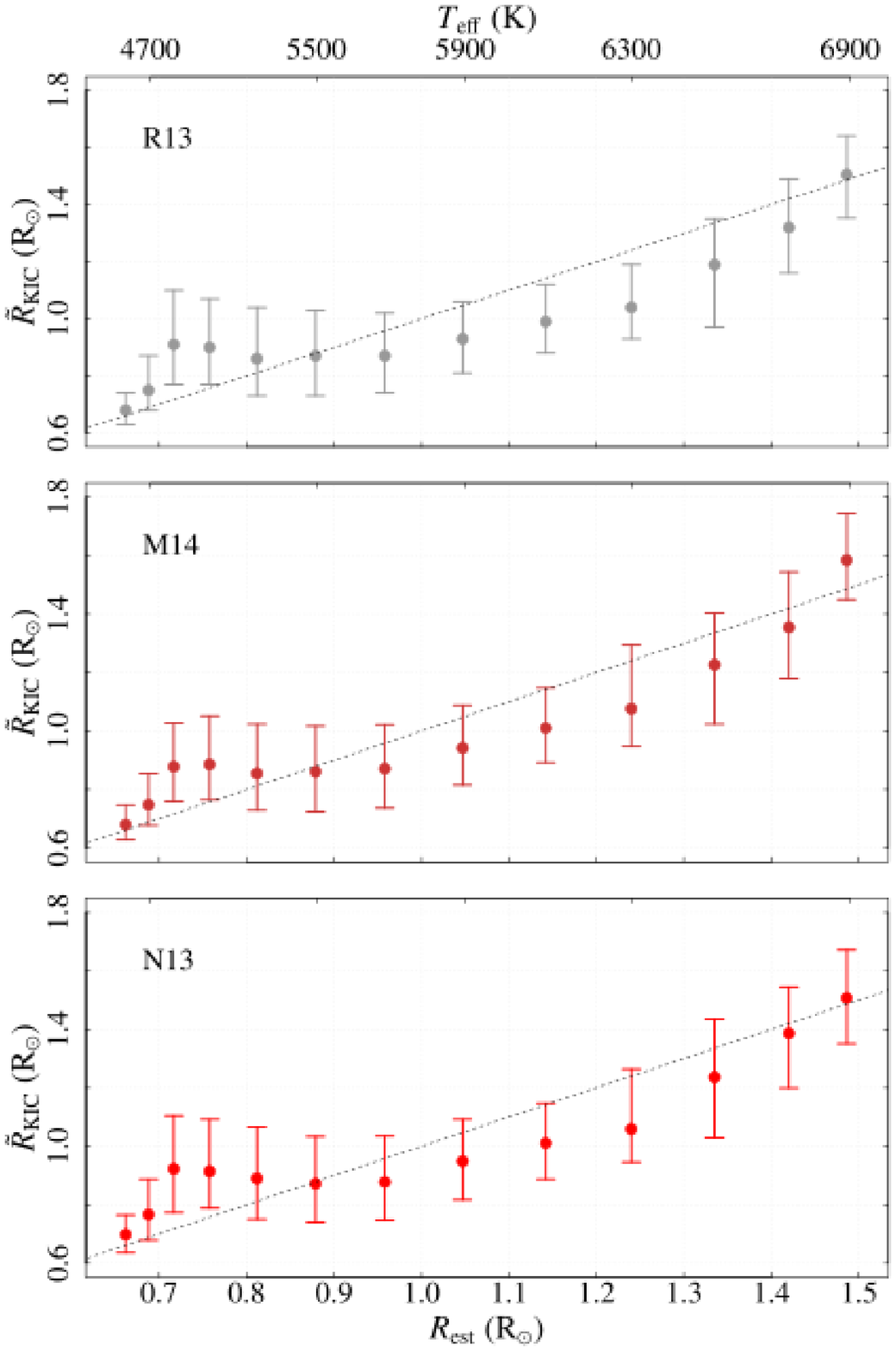}
  \caption{Median KIC radii, $\mrkic$, as a function of estimated typical radii, $\rest$, for stars in
each $\tef$ bin (see Table \ref{data}). The error bars are the  first and third quartiles of the
distribution of $\rkic$ in the $\tef$ bin. The black curve is the line of equality, and the top axis
corresponds to the effective temperature.}
   \label{radii}
 \end{figure}

Radii from KIC were estimated from $\log(g)$ using a mass--radius relationship. They have
uncertainties due to errors in photometry, degeneracies between stellar parameters and colors, as well as
errors from stellar models \citep[cf.][]{Gaidos13}. \citet{Verner11} compared KIC and astroseismic radii for
500 solar-type stars and found an underestimation bias of up to 50\% for stars with $\rkic< 2\rsun$. Such a
discrepancy is the result of the overestimated values of the KIC $\log(g)$. Analyzing uncertainties 
in the KIC radii of KOI stars, \citet{Gaidos13} found that K-type dwarfs have inaccuracies less than 
about 15\%, while F- and many G-type stars have uncertainties that can be higher than 100\%. We 
estimated the typical radii for the stars in our samples using the mean temperature of the stars in 
each $\tef$ bin considered (see Table 1), according to the calibration between $\tef$ and radii 
given in Table B1 by \citet{Gray92}. The estimated radii, $\rest$, as well as the median KIC radii, 
$\mrkic$, is given in Table \ref{data}. Figure \ref{radii} shows that $\mrkic$ are overestimated 
relatively to $\rest$ for stars with $\tef\lesssim5\,500$\,K, and underestimated for stars with 
$\tef$ in the range $5\,500\lesssim\tef\lesssim6\,900$\,K. However, there is a reasonable agreement 
between $\rkic$ and $\rest$, by considering the error bars. The mean relative difference between 
$\rkic$ and $\rest$, relatively to $\rest$, is around 10\% for all the samples. The effect of an 
overestimation (or underestimation) of the median radius is a systematic proportional increase (or 
decrease) in $\mv$. It is clear that $\rest$ is a rough approximation of the central tendency of 
stellar radii in each $\tef$ interval. In fact, it does not consider many factors influencing the 
radius estimates, such as the stellar composition and evolution in the main sequence. The whole 
point of comparing $\mrkic$ and $\rest$ is to show that the lower and higher $\mrkic$ presented in 
Table \ref{data} are reasonable estimates of the radius variation limits for the stars in our 
samples, despite the large uncertainties in $\rkic$.

In Figure \ref{means} (panels (b), (d), and (f)), we present the behavior of $\momega$ as a 
function of $\mvsini$. The cyan region corresponds to the angular rotations estimated from 
$\mvsini$, according to the equation $\momegam=(\pi/4)^{-1}R^{-1}\mvsini$. For each sample, the 
parameter $R$ ranges from the minimum to the maximum $\mrkic$ given in Table \ref{data}. Panel 
(b) reinforces the statement that there is a very good agreement between the behavior of the $\mv$ 
as a function of $\mvsini$ and the CR model. In fact, if we consider that the radii of the stars 
range approximately from the minimum to the maximum $\mrkic$, we have that 12 in 13 groups support 
the CR model with a 95\% confidence level. It is also worth noting that even a systematic 
decrease of $4\kms$ in $\vsini$ is sufficient to bring the outside group into the cyan region. Such
a systematic shift down is reasonable because the $\vsini$ data from N04 are biased to up 
($5$-$10\kms$) to values higher than $30\kms$, as we mentioned in Section \ref{vsini}.

In the panel (d), we observe that 11 in 13 groups of stars present a behavior of $\momega$ with
$\mvsini$ consistent with the CR model. The groups outside the cyan area have $\mvsini>30\kms$. In order to
make these two groups agree to the CR model, it is required that a systematic decrease of at least 
13, and $23\kms$ in $\vsini$, or equivalently, a systematic decrease of about 30, and 45\% in 
$\prot$. Panel (f) shows that the groups from the N13 sample with $\vsini\gtrsim15\kms$ seem to rule 
out the model. For the groups with $15\lesssim\mvsini<30\kms$, a systematic shift down to about 
$2\kms$ in $\vsini$, or an increase of $\sim10$\% in $\prot$ is necessary for these groups to agree 
with the model. The groups presenting the larger discrepancies contain high rotators, with 
$\vsini>30\kms$. To correct these discrepancies, it is necessary for a systematic reduction of at 
least 19, and $23\kms$ in $\vsini$, or a decrease of around 45\% and 44\% in the rotation periods. 
The large discrepancies between the CR model and the data for the samples M14 and N13 cannot be 
explained by the uncertainties in $\vsini$ data. As we mentioned in Section \ref{worksample}, the 
uncertainties in $\vsini$ measurements are $1\kms$, for the stars with $\vsini<30\kms$, and 
$5$-$10\kms$, for the stars with $\vsini<30\kms$. However, these apparent divergences from the CR 
model presented by the samples M14 and N13 seem to be in part due to the selection effects that we 
will discuss bellow.

Distinguishing between the low rotation period from stellar activity and the period from other 
sources such as stellar pulsations is a difficult task. A lower limit was imposed for $\prot$ in 
the original samples to try to work around this problem. The lower limit period for the 
samples R13, M14, and N13 are 0.5, 0.2, and 1\,day, respectively 
\citep{Reinhold13,Nielsen13,McQuillan14}. As a consequence, there is a bias in these samples 
against the higher rotators, because the stars with $\prot$ below the lower limit are left out. 
These biases are most important in the N13 sample, as its lower limit is one day, and in the 
M14 sample, which has an additional bias due to the upper limit imposed on the effective 
temperature. This bias results in an underestimate of the true $\mv$ and may be an important factor 
in the discrepancies between the CR model and $\mv$ observed for the higher rotators pictured in 
Figure \ref{means}.

In order to overcome this observational bias against the high rotators, we estimated the mean 
rotations using the nonparametric maximum likelihood estimator (NPMLE) of the distribution 
function of $V=50.58\rkic/\prot$ with right truncation $V\lesssim 50.58\rkic/P_{\rm min}$, where 
$P_{\rm min}$ is the lower limiting period of the sample. The NPMLE was computed using the algorithm 
proposed by \citet{EfronPet99}, which use the \citet{LynBell71} method \citep[see][]{Moreiraetal10}. 
The corrected means for the sample R13 are $93^{+44}_{-40}\kms$, and $81^{+10}_{-9}\kms$. For the 
M14 sample these means are near $35^{+1}_{-1}\kms$, and $48^{+0}_{-4}\kms$.  The N13 sample has 
corrected means $102^{+21}_{-61}\kms$, and $121^{+22}_{-65}\kms$. These means are presented in 
Figure \ref{means} (panels (a), (c), and (e)) in green. The error bars were estimated using 
the 95\% confidence bands of the NPMLE estimated through bootstrap resamples with 1000 replacements. 
As can be observed, the corrected means are higher than the bootstrapped means $\mv$, 
reinforcing the idea that a bias against the higher rotators contributed to the discrepancies 
between the CR model and $\mv$ observed for the stars with $\vsini>30\kms$. The difference is more 
significant in the N13 sample, where the minimum rotation period is 1\,day. For the M14 sample, 
where the minimum rotation period is only 0.2\,day, the corrected means do not differ 
significantly from the bootstrapped means. However, we must take into account that there is an 
additional bias against the high rotators in M14. As mentioned in Section \ref{periods}, 
\citet{McQuillan14} removed the stars with $\tef\lesssim 6500$\,K from their sample, which is 
equivalent to $\tef\lesssim 6755\pm67$, in \citet{Pinsonneault12} temperature scale.\footnote{To 
convert the temperature from the scale adopted by \citet{McQuillan14} into the scale adopted in the 
present work, we used the calibration $\tef[{\rm Pin}]=(1.05\times \tef[{\rm Mc}] - 76.32)$K, with 
rms $=67.4$\,K. This calibration is a linear regression between the temperatures $\tef$ calculated 
by \citet{McQuillan14} ($\tef[{\rm Mc}]$), and $\tef$ calculated using the \citet{Pinsonneault12} 
($\tef[{\rm Pin}]$) for the same stars.} This procedure also excludes stars with higher rotations.

On the other hand, the objects in the original samples were selected considering an upper limit of 
the period, which varies depending on the method used to calculate $\prot$ (see Section 
\ref{periods}). This introduced a bias in our samples against the lower rotators. The main effect of 
this bias is an overestimation of the lowest means $\mv$. As we can see in Figures \ref{density} and 
\ref{means}, such a bias is more pronounced in N13 because the method of determining period by 
\citet{Nielsen13} requires $\prot<30$\,days. With the objective of testing the consistency of their 
measurement with $\vsini$ data, \citet{Nielsen13} compared $\mv$, with $\mvsini$, estimated applying 
CR to $\vsini$ data from the compilation by \citet{Glebocki05}. They found a good consistency 
between these median rotations for the stars in the spectral region F0-K0 
($7\,000\lesssim\tef\lesssim 5\,000$\,K), and an inconsistency for the later type stars. Such an 
inconsistency was linked to the presence of young open cluster stars in the \citet{Glebocki05} 
compilation. However, the result seems to show that these differences are most likely due to 
selection effects in their sample, at least in the range of $\tef$ we analyzed.

Finally, it is worth considering the selection effect associated to the age--activity relation. 
The rotation periods are based on the modulations in the stellar LCs, which in, turn are due 
to changes in the stellar surface caused by stellar activity. Since both rotation and activity 
decrease with age \citep[see][]{Vaiana92, Covey11}, this may cause older and less active stars 
(slowest rotators) to be measured more rarely than the younger and more active ones (fastest 
rotators), which can result in a super-estimation of the actual mean rotation $\mv$. Aiming to have 
an overall view on the influence of this bias in our results, we analyzed subsamples drawn from N04, 
R13, M14, and N13 according to the stellar age: $<4.5$\,Gyr, $<3$\,Gyr, and $<2$\,Gyr. The age for 
an individual star in N04 is given in \citet{Holmberg09}. The maximum age of the 
stars in samples R13, M14 and N13 were estimated using the gyrochronology relations by 
\citet{Mamajek08}. The subsamples aged less than 2\,Gyr is shown in Figure \ref{density}. We also 
imposed a low limit to the effective temperature as $\tef=4\,900$\,K because below that temperature 
there is a lack of stars in the subsample of $\vsini$ data. With the purpose of illustrating this 
behavior, Figure \ref{means} (panels (a), (c), and (e)) shows the best-fit line (in blue) for the 
subsamples with age $<2$\,Gyr. The analysis of the behavior of $\mv$ as a function of $\mvsini$ 
shown that the best-fit line for the data tends to the CR as we reduced the maximum age of the 
subsample. This analysis seems to show us that the agreement between the CR and the data is improved 
by minimizing the bias due to the age--activity relation despite it not completely solving the 
problem of bias.

\section{Conclusions}
\label{conclusions}

The methods of measuring projected rotation from the spectral profile allowed a large amount of 
$\vsini$ data in the literature. These observational data are commonly used to estimate the mean of 
true rotation for a group of stars using the CR. Unfortunately, the lack of large and homogeneous 
sample of field stars while taking measured $\vsini$ and rotation period does not allow a general 
and conclusive test of this relation. The purpose of this study was to conduct a statistical test of 
the CR and the hypothesis of the constant mean $\msini=\pi/4$ for the stars in the {\it Kepler} 
field. This study was carried out using three homogeneous samples of rotation periods of stars in 
the {\it Kepler} field and a homogeneous sample of $\vsini$ in the Galactic field. The samples were 
segregated by the effective temperature interval and the mean rotations as well as the 95\% 
confidence intervals of these means, were estimated using the bootstrap-resampling method.

In summary, we found that the distribution of the mean true rotation as a function of the mean projected
rotation present a very good agreement to the CR, with $\msini=\pi/4$, for one of the
samples, and present some discrepancies for the other ones. Such discrepancies may be 
due to bias arising from the uncertainties in the KIC radii, and selection effects. Our results 
pointing out the statistical consistence between the behavior of the observational 
data and the CR with the constant $\msini=\pi/4$, reinforce the CR as an appropriate way to 
estimate $\mv$ from $\mvsini$ for populations of field stars. We consider that the main 
contribution of this paper is to reinforce the statement that there is no preferential 
orientation of the stellar rotation axes in the Galactic field, and that the CR with $\msini=\pi/4$ 
can constitute a key test to new methods for estimating rotation periods by using $\vsini$ data.

\acknowledgments
We thank the anonymous referees for providing constructive comments that strengthened the 
manuscript and the reported results. We are also grateful to Dr. J.-D. do Nascimento Jr., for 
discussing many of the issues in this paper.


\end{document}